\documentclass[10pt,
twocolumn,
showpacs,
amsmath,amssymb,
aps,
prb,
floatfix,
]{revtex4-1}
\usepackage{amsmath}
\usepackage{graphicx}
\usepackage{amssymb}
\usepackage{color}

\newcommand{\comm}[1]{\ensuremath{\left[#1\right]_-}}

\newcommand{\fn}{\ensuremath{\mathbf{n}}}
\newcommand{\fs}{\ensuremath{\mathbf{s}}}
\newcommand{\fJ}{\ensuremath{\mathbf{J}}}
\newcommand{\dsd}{\ensuremath{\Delta_{\text{sd}}}}

\newcommand{\dra}{\ensuremath{\Delta_{\text{R}}}}

\newcommand{\fJp}{\ensuremath{\mathbf{J}_p}}

\newcommand{\fey}{\ensuremath{\mathbf e_{y}}}

\newcommand{\xdw}{x_{\rm{DW}}}
\newcommand{\vdw}{v_{\rm{DW}}}

\newcommand{\fmp}{\ensuremath{\mathbf m_p}}

\newcommand{\half}{\ensuremath{\frac{1}{2}}}

\newcommand{\mean}[1]{\ensuremath{\left\langle#1\right\rangle}}

\newcommand{\dx}{\ensuremath{\partial_x}}
\newcommand{\dt}{\ensuremath{\partial_t}}

\begin{document}
\title{Rashba induced chirality switching of domain walls and suppression of the
Walker breakdown}

\author{Martin Stier, Marcus Creutzburg, and Michael Thorwart}
\affiliation{I. Institut f\"ur Theoretische Physik, Universit\"at Hamburg, 
Jungiusstra{\ss}e 9, 20355 Hamburg, Germany}

\begin{abstract}
We investigate the current-induced motion of ferromagnetic domain walls in
presence of a Rashba spin-orbit interaction of the itinerant electrons. We show
how a Rashba interaction can stabilize the domain wall motion, such that 
the Walker breakdown is shifted to larger current densities. The Rashba
spin-orbit interaction creates a field-like contribution to the spin torque,
which breaks the symmetry of the system and modifies the internal structure of
the domain wall. Moreover, it can induce an additional switching of
the chirality of the domain wall for sufficiently strong Rashba interactions. 
This allows one to choose the desired chirality by the choosing the direction
of the applied spin-polarized current. Both the suppression of the Walker
breakdown and the chirality switching affect the domain wall velocity
significantly. This is even more pronounced for short current pulses, where an
additional domain wall drift  in either positive or negative
direction appears after the pulse ends. By this, we can steer the final position
of the domain wall. This mechanism may help to
overcome the current limitations of the domain wall motion due to the Walker
breakdown which occurs for rather low current densities in systems without a
Rashba spin-orbit interaction. 
\end{abstract}


\maketitle

\section{Introduction}

Magnetic memory devices are based on the presence of microscopic magnetic
domains where the alignment, or more precisely the alternation of alignments, of
the magnetization encodes the physical information. While in classical hard
discs, these domains are directly switched by a magnetic field, it is
energetically far more efficient to push them through a wire by an electrical
current\cite{parkin2008magnetic,malinowski2011current,brataas2012current,
fert2013skyrmions}. The central feature of this current-induced domain wall (DW)
motion can be rationalized within the picture of the standard $sd$-model of the
localized electrons which form the local magnetic moments and the itinerant
electrons injected into the $s$-band whose spins are polarized from outside in
external contacts \cite{Zener,berger1985}. Then, the local exchange interaction
of the itinerant electron spins with the local magnetic moments generates  
 a current-induced spin-transfer torque (STT). Two basic contributions of the
STT are well known in materials without a breaking of the symmetry in the
 subsystem of the itinerant
electrons\cite{zhang2004roles,brataas2012current}. The adiabatic STT stems from
the adiabatic alignment of the itinerant spins and the local magnetic moments,
resulting in a DW motion due to the conservation of the total spin. Moreover,
the nonadiabatic STT arises, which is also known as the $\beta$-term. It has
its origin in a lag of the dynamics of the polarization of the itinerant
electrons behind the dynamics of the local magnetic texture. This back-action
of the local magnetic moments on the itinerant spins induces a relaxation
dynamics for the latter during which an additional nonadiabatic current-induced
STT is generated. Even though the
non-adiabatic STT can lead to a considerable increase of the DW velocity with
respect to the adiabatic motion, a very fast movement of the DW is limited by
the Walker breakdown (WB) at a critical current density, which is accompanied by
a precession of the magnetization\cite{hayashi2006direct} at the DW center.
However, this precession can be suppressed in systems with a broken symmetry
where a distinct direction of the magnetization or the electron spin is favored
over the others. This may also imply a preference for a distinct chirality, or
handedness, of the DW. Based on this observation, Miron \textit{et
al.}\cite{miron2011} have proposed to use the Rashba effect as a stabilizer of
the DW chirality and the corresponding suppression of the WB and an increase of
the DW velocity have been observed experimentally
\cite{glathe2008experimental,obata2008current}. The mechanism is similar to the
action of a transverse magnetic
field. In addition, other mechanisms to break the symmetry, such as
the Dzyaloshinskii-Moriya interaction, are supposed to enhance the DW motion due
to a preferred
handedness\cite{brataas2013spintronics,ryu2013chiral,emori2013current}. Hence,
by modifying the thickness of layers of distinct materials, the strength of the
Dzyaloshinskii-Moriya interaction, and with it, the preferred chirality, can
even be adjusted within certain limits\cite{chen2013tailoring}.

The aim of this work is to reveal how a Rashba spin-orbit interaction of the
itinerant electron spins acts on the chirality and the dynamics of a DW. We
thereby consider the stabilization or destabilization of a distinct chirality in
certain parameter regimes and their impact on the DW velocity. To illustrate
the basic physical mechanism at work, we use a one-dimensional (1D) model which
allows us to calculate the full STT including the Rashba-induced effective
field in simple terms. By eventually solving the Landau-Lifshitz-Gilbert
equation of motion of the DW and hereby calculating DW velocities for opposite
chiralities, we identify several regimes of chirality-dependent DW motion.
This actually includes a regime, where WB is suppressed and shifted to larger
current densities, but also a current-dependent switching to the
desired chirality. We show that the optimal chirality can be chosen by the
direction of the applied current flow. Results are presented in a broad
parameter range of the current density, the strength of the Rashba
interaction and the lengths of the applied current pulse. Particularly
for short current pulses, we find that the the average DW velocity may differ
strongly from the steady current value.

\section{Model}

We consider a 1D quantum wire in which a DW is formed by localized magnetic
moments
$M_s\fn(x,t)$ described by a unit vector $\fn(x,t)$ and its saturation
magnetization $M_s$. The dynamics of these classical moments is well
described by the Landau-Lifshitz-Gilbert
equation of motion \cite{lakshmanan2011fascinating}.
It describes the precessional motion of the moments which can be either
induced by external global magnetic fields or local interactions. This
precession is damped by the Gilbert damping term which lets the moments to
actually align towards the (local or global) magnetic field. In our case the
Landau-Lifshitz-Gilbert equation 
\begin{equation}
 \dt\fn = -\gamma_0\fn\times\mathbf H_{\rm{eff}} + \alpha \fn\times\dt\fn +\mathbf
T \label{eq::llg}\ ,
\end{equation}
includes an effective field $\mathbf H_{\rm{eff}}$, the gyro-magnetic
ratio $\gamma_0$, the Gilbert damping constant $\alpha$ and a 
spin torque $\mathbf T$ which is provided by external means and which is in the
focus of the present work. In this work, spin transfer torque (STT) $\mathbf T$
has its origin in a spin polarized current which interacts with the magnetic
moments via the exchange interaction. 
 
The actual shape of the DW is created by an effective magnetic field
 which represents the interactions between the magnetic moments. We use the
standard continuum form\cite{lakshmanan2011fascinating,li2004} 
\begin{equation}
 \mathbf H_{\rm{eff}}=-J_{\rm{IA}}\dx^2\fn -K_{\parallel} n_{\parallel} \mathbf
e_{\parallel}+K_{\perp} n_{\perp} \mathbf e_{\perp}\ .
\end{equation}
Here, the interaction (IA) strength $J_{\rm{IA}}=2A_{\rm ex}/M_s$ stems from the
mutual exchange interaction among the spins with the strength $A_{\rm ex}$ and 
is a measure for the tendency of the magnetic moments to align parallel to each
other. Moreover, the easy-axis anisotropy $K_{\parallel}$ and the hard axis
anisotropy $K_{\perp}$ are the energies of the favorable and unfavorable
directions for these moments, respectively. The explicit directions of these
axes (e.g., the $x,y$ or $z$ direction)
have to be defined according to the situation under consideration and different
constellations arise. Even though the choice of the hard and the easy axis is
hardly of any importance in highly symmetric systems, it becomes relevant if the
symmetry is broken. Since this is the case for systems with a Rashba
spin-orbit interaction in the focus here, we will consider different setups in
this work.

A major point in this work is the calculation of $\mathbf T$ in Eq.
(\ref{eq::llg}) in presence of a Rashba spin-orbit interaction and a
nonadiabatic relaxation channel for the itinerant electron spins. To do this, we
make use of the standard $sd$-model \cite{Zener, berger1985} of the electrons in
the quantum wire and describe the localized magnetic moments $\fn(x,t)$ as the
spins of the localized electrons which typically live in $d$-like bands. They
couple to the spins $\fs(x,t)$ of the flowing or itinerant electrons in the 1D
wire which typically live in $s$-like bands. Those are assumed to be
non-interacting and are described by the kinetic Hamiltonian $H_{\rm{kin}}$. 
The two species couple via the
exchange or the $sd$-interaction which gives rise to the Hamiltonian 
$H_{\rm{sd}}$. In addition, we
add to this minimal 1D model the Rashba spin-orbit interaction
$H_{\rm{Rashba}}$ for the itinerant electron spins in the spin polarized
electron current which is imprinted at the ends of the quantum wire (see
below). Since we are interested in nonadiabatic effects, we include a
relaxational part $H_{\rm{relax}}$ for the itinerant electrons. In total, this
yields the Hamiltonian 
\begin{equation}
 H = H_{\rm{kin}} + H_{\rm{sd}} + H_{\rm{Rashba}} + H_{\rm{relax}}\
.\label{eq::totham}
\end{equation}
\begin{figure}[tb]
\centering
 \includegraphics[width=.9\linewidth]{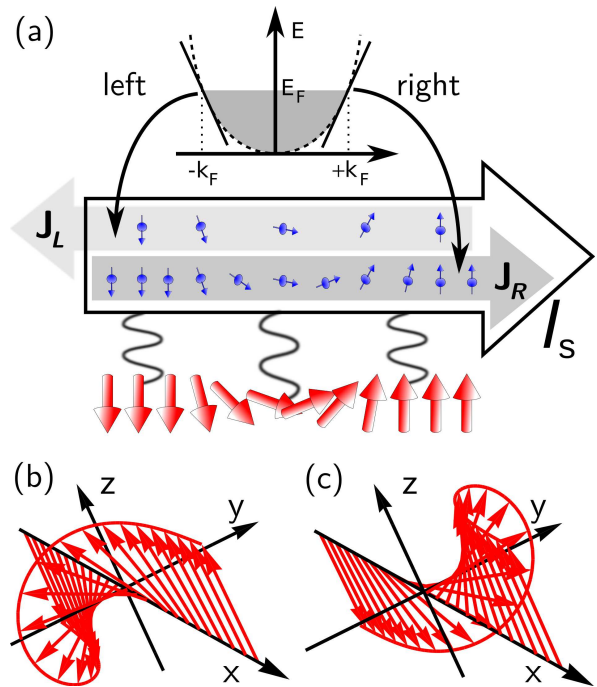}
 \caption{\label{fig::scheme}(color online) (a) Schematic view on the
linearization of the original electron dispersion (dashed) to two branches of
left/right moving particles (full lines) in the vicinity of the Fermi energy and
the Fermi wave vector $\pm k_F$. This leads to left or right moving spin
(polarized) currents which form a total spin current with a density $I_s$. Both,
left and right moving electrons (blue), couple to the DW's local moments (red)
via the $sd$ interaction. (b/c) Schematic view of a Bloch(z) DW with hard $x$
anisotropy and positive chirality (b) and negative
chirality (c).}
\end{figure}

We are only interested in 1D systems and it is very convenient to use the 1D
Sugawara representation of the Hamiltonian of the spin sector. It is 
an  appropriate description for 1D systems\cite{haldane1981luttinger,
gogolin2004bosonization} and provides a rather simple and straightforward way to
calculate the STT \cite{thorwart2007,stier2013}. In the Sugawara
representation, we concentrate on the low energy sector of the electronic
system. This means that we only focus on excitations in the vicinity of the
Fermi level. As the dispersion is only varying slowly in this region we can
linearize the dispersion which yields two chiral branches of the dispersion (cf.
Fig. \ref{fig::scheme}a). These branches can be associated to left and right
moving electrons in the 1D wire. This  yields the standard form of
the kinetic part of the low energy Hamiltonian for both spin directions
\begin{equation}
 H_{\rm{kin}} = -i\hbar v\sum_{\sigma,p}\int dx c_{p\sigma}^{\dagger}(x) \dx
c_{p\sigma}(x),
\end{equation}
where $c_{p\sigma}^{(\dagger)}$ are the annihilators (creators) of electrons
with spin $\sigma=\uparrow,\downarrow$ which are moving in the left or right
direction ($p=L/R=-/+$) and the Fermi velocity $v$. To rewrite this in the
Sugawara form, we define the spin density
operators\cite{gogolin2004bosonization} 
\begin{equation}
 \fJp(x) = \half :c_{p\sigma}^{\dagger}(x) \boldsymbol\sigma_{\sigma\sigma'}
c_{p\sigma'}(x):
\end{equation}
with the Pauli matrices $\boldsymbol\sigma$ and the colons $:\dots :$ denoting
normal ordering. The Hamiltonian now reads
\begin{equation}
  H_{\rm{kin}} = \hbar v \sum_p \int dx\ :\fJp\cdot\fJp: + H_{\rm{charge}}
\end{equation}
with an irrelevant charge part. As we set-up the equation of
motion for $\fJp$ below and $\comm{\fJp,H_{\rm{charge}}}=0$, the charge part
does not contribute to the equation of motion. On the basis of the spin
density operators for left and right moving particles, we find  a
simple definition of the total spin density
\begin{equation} \label{eq::spindens}
 \fs = \fJ_R+\fJ_R \, ,
\end{equation}
and, more importantly, of the spin current density
\begin{equation}
 \fJ = v(\fJ_R-\fJ_L)\ ,\label{eq::spincurrdef}
\end{equation}
which reduces to a vector instead of a tensor in 1D. All remaining parts of the
total Hamiltonian (\ref{eq::totham}) can also be expressed in terms of the
$\fJp$. This is obvious for the $sd$ Hamiltonian which gives 
\begin{equation}
 H_{\rm{sd}}=\dsd\int dx\ \fs\cdot\fn=\dsd\sum_p\int dx\ \fJp\cdot\fn\ .
\end{equation}
Regarding the Rashba Hamiltonian $H_{\rm{Rashba}}$, we can use further
simplifications which arise in 1D wires. The conventional Rashba Hamiltonian
\begin{equation}
 H_{\rm{Rashba}} = \tilde\alpha_R(k_x\sigma_y - k_y\sigma_x)
\end{equation}
reflects motion in 2D, while we may neglect the movement in one of the
directions in 1D wires. Thus, we only keep $H_{\rm{Rashba}} = \tilde\alpha_R
k_x\sigma_y$ for our calculations and ignore some higher-order admixing of
transverse states\cite{schulz2009low}. Additionally, in a low-energy model only
wave vectors
in the vicinity of the Fermi wave vector $k_F$ are relevant and we can replace
$k_x\to k_F$ for the right moving and $k_x\to-k_F$ for the left moving
particles. This yields the simplified 1D Rashba Hamiltonian in the Sugawara
form 
\begin{equation}
 H_{\rm{Rashba}} = \dra\sum_p p\int dx\ \fJp\cdot \mathbf e_y,\quad
\dra=2\tilde\alpha_R k_F.
\end{equation}
For simplicity, we combine the two interactions to
\begin{align}
 H_{\rm{IA}} \equiv& H_{\rm{sd}}+H_{\rm{Rashba}}\nonumber\\
 =& \dsd\sum_p\int dx\ \fJp\cdot\fmp\label{eq::IA}\\
 \fmp =& \fn+p\alpha_R \fey\label{eq::mp}
\end{align}
and introduce a reduced Rashba interaction $\alpha_R = \dra/\dsd$. This also
illustrates that the Rashba interaction provides an effective local magnetic
field for the localized magnetic moments $\fn(x,t)$ which depends on the chiral
index $p=L/R$. 

The last part of the total Hamiltonian (\ref{eq::totham}) is the relaxation
part, which we define here implicitly by the help of the commutator 
\begin{equation}
-\frac{i}{\hbar}
\comm{\fJp,H_{\rm{relax}}}=\frac{1}{\tau}(\fJp-\fJp^{\rm{relax}})\ .
\end{equation}
This form results from a standard relaxation time approximation with the
relaxation time $\tau$ and can be derived from a microscopic system-bath
Hamiltonian on the basis of a Bloch-Redfield-like approach. We refer
to Ref. \onlinecite{thorwart2007} for further details.

\subsection{Equation of motion for itinerant electrons}

Next, we formulate the Heisenberg equation of motion (EOM) of $\fJp$ as 
\begin{equation}
 \dt\fJp=-\frac{i}{\hbar}\comm{\fJp,H}\ .
\end{equation}
Its solution enters in Eq.\ (\ref{eq::spindens}) and eventually yields the STT
\begin{equation}
 \mathbf T =
-\frac{\dsd}{\hbar}\fn\times\fs=-\frac{\dsd}{\hbar}\fn\times\sum_p\fJp\
.\label{eq::stt}
\end{equation}
Keeping in mind that the spin density operators in the low-energy description
obey the Kac-Moody-Algebra\cite{gogolin2004bosonization} with the commutators 
\begin{equation}
 \comm{J_p^{\mu}(x),J_{p'}^{\nu}(x')}=i[p\dx+\epsilon^{\mu\nu\lambda}J_p^{
\lambda}]\delta_{pp'}\delta(x-x'),
\end{equation}
we find
\begin{align}
 (\dt+vp\dx)\fJp
=&-\frac{\dsd}{\hbar}[\fJp\times\fmp+\beta(\fJp-\fJp^{\rm{relax}})],
\label{eq::eom}
\end{align}
where we have defined $\beta=\hbar/(\dsd\tau)$. The explicit form of the
relaxation state is crucial for the resulting STT, since it not only changes the
values of the equation but also influences the symmetry of the system. As 
shown by van der Bijl and Duine \cite{vanderbijl2012}, this actually
affects the existence of distinct STTs. To actually solve the EOM
(\ref{eq::eom}), we apply a gradient expansion scheme and express $\fJp$ in
orders of derivatives of $\fmp$ as
\begin{equation}
 \fJp = \fJp^{(0)}(\fmp) + \fJp^{(1)}(\dx\fmp,\dt\fmp)+\dots \
.\label{eq::gradient}
\end{equation}
Notice that obviously $\partial_{x,t}\fmp = \partial_{x,t}\fn$. The combination
of
the Ansatz (\ref{eq::gradient}) and the EOM (\ref{eq::eom}) allows for an
arrangement by the orders of the derivatives on the respective left and right
hand side of the equation as
\begin{align*}
 0=&-\dsd[\fJp^{(0)}\times\fmp-\beta(\fJp^{(0)}-\fJp^{\rm{relax}})]\\
  (\dt+vp\dx)\fJp^{(0)} =&-\dsd[\fJp^{(1)}\times\fmp-\beta(\fJp^{(1)}- 0 )]\\
  \dots=&\dots\ .
\end{align*}
Every equation has the basic structure 
\begin{equation}
 -\dsd(\beta -\fmp\times)\fJp^{(n)} = \mathbf X_p^{(n)}
\end{equation}
with $\mathbf X_p^{(0)}=-\dsd\beta\fJp^{\rm{relax}}$ and $\mathbf
X_p^{(1)}= (\dt+vp\dx)\fJp^{(0)}$ and so on. The general solution
reads\cite{thorwart2007}
\begin{equation}
 \fJp^{(n)}=\frac{\beta^2\mathbf X_p-\beta\mathbf X_p^{(n)}\times\fmp+(\mathbf
X_p^{(n)}\cdot\fmp)\fmp}{\beta(\beta^2+\fmp^2)}\ .\label{eq::gen_sol}
\end{equation}
Starting from the zeroth order, we can now solve the equation successively to,
in
principle, arbitrary order. Since higher order terms become very involved, we
restrict the calculation to zeroth and first order in this work.

Next, we have to address the relaxation state explicitly. In the literature,
two approaches are discussed: the electron spin either relaxes towards the
direction of the
magnetization\cite{kim2012magnetization}, such
that $\fJp^{\rm{relax}}\propto\fn$, or to
the combined vector\cite{vanderbijl2012} $\fJp^{\rm{relax}}\propto\fmp$.
To reveal the differences between these two approaches, we address below 
both.

\subsection{\label{sec::stt}Spin torque}

\subsubsection{Relaxation to $\fn$}

In the first approach, the relaxation occurs towards the magnetization $\fn$ of
the domain wall, such that 
\begin{equation}
 \fJp^{\rm{relax}}=j_p \fn
\end{equation}
with some proportionality coefficients $j_{L/R}$. Considering small damping
parameters $\beta^2\ll 1$, we find for the zeroth order
\begin{equation}
 \fJp^{(0)}=j^{(0)}_p\frac{-\beta\fn\times\fmp}{(\fn\cdot\fmp)|\fmp|}+j_p^{(0)}
\frac{\fmp}{|\fmp|}\ .\label{eq::jp0_rel2n}
\end{equation}
Here, we have introduced $j_p^{(0)}=j_p(\fn\cdot\fmp)/|\fmp|$ to ensure that 
$|\fJp^{(0)}|\stackrel{\beta^2\ll 1}{=}j_p^ {(0)}$. By this, the
spin current density far away from any magnetic texture, i.e., for $x\to \pm
\infty$ in the zeroth order follows from Eq. (\ref{eq::spincurrdef}) as 
\begin{equation}
 I_s\equiv v |\fJ^{(0)}_R-\fJ^{(0)}_L|\stackrel{\alpha_R^2,\beta^2\ll
1}{=}v(j_R^{(0)}-j_L^{(0)}) \, .\label{eq::is}
\end{equation}
For an easier comparison with experimental data, this spin current density may
be rewritten as $I_s=PI_c/(2eM_s)$, where $P$ is the spin polarity, $e$ the
elementary charge, $M_s$ the saturated magnetic moment, and $I_c$ the charge
current density of the imprinted spin polarized current. Using Eqs.
(\ref{eq::stt}) and (\ref{eq::is}), we
find the zeroth order contribution to the STT as 
\begin{equation}
 \mathbf T^{(0)}=-\frac{\dsd}{\hbar
v}I_s\alpha_R\Big[\fn\times\fey-\beta\fn\times(\fn\times\fey)\Big]+\mathcal
O(\alpha_R^3)\ .
\end{equation}
This equation may be expressed in the form of a term with an effective magnetic
field as it appears in the LLG (\ref{eq::llg}), such that 
\begin{equation}
  \mathbf T^{(0)} = -\gamma_0\fn\times(\mathbf H_R^{(0)} + \mathbf
H_R^{\rm{anti}}) \, .
\end{equation}
This form gives rise to the ``Rashba field''
\begin{equation}
  \mathbf H_R^{(0)}= \frac{\dsd}{\hbar v\gamma_0}I_s\alpha_R\fey \, ,
\end{equation}
and to the ``anti-damping field ''
\begin{equation}
  \mathbf H_R^{\rm{anti}}= -\frac{\dsd}{\hbar v\gamma_0}\beta
I_s\alpha_R\fn\times\fey \, .
\end{equation}
From the zeroth order spin density (\ref{eq::jp0_rel2n}), we obtain the first
order term which eventually yields the usual adiabatic and nonadiabatic
contributions to the STT as 
\begin{align}
 \mathbf T^{\rm{ad}} =& -I_s\dx\fn+\mathcal O(\alpha_R^2)\\
 \mathbf T^{\rm{non-ad}} =& \beta I_s\fn\times\dx\fn+\mathcal O(\alpha_R^2)\, ,
\end{align}
as well as a first order contribution to the Rashba field as
\begin{equation}
   \mathbf H_R^{(1)}=
-\frac{I_s}{\gamma_0}\alpha_R^2\big[(\fn\times\dx\fn)\cdot\fey\big]\fey+\mathcal
O(\alpha_R^3)\ .
\end{equation}
When the term $(\fn\times\dx\fn)\cdot\fey$ is large, $\mathbf H_R^{(1)}$ may
strongly affect the whole Rashba field, even though it is proportional to
$\alpha_R^2$. However, for Bloch-like DWs this term appears to be small and we
will not focus on it in this work.

Additional terms $\mathbf T_t\propto\dt\fn$ also appear, which renormalize the
Gilbert damping $\alpha$ in the LLG. As the origin of $\alpha$, and with it, the
dependence
on other parameters is not very well established in theory, we will neglect all
torques $\propto\dt\fn$ to obtain a constant model parameter $\alpha$ for all
calculations shown below. 

\subsubsection{Relaxation to $\fmp$}

When the relaxation state is chosen as  
\begin{equation}
 \fJp^{\rm{relax}}=j_p\fmp \, ,
\end{equation}
the zeroth-order solution is 
\begin{equation}
 \fJp^{(0)}=j_p^{(0)}\frac{\fmp}{|\fmp|}\ ,
\end{equation}
with $j_p^{(0)}=j_p|\fmp|$. Consequently, the anti-damping term is missing in
the zeroth order contribution of the STT
\begin{equation}
 \mathbf T^{(0)}=-\gamma_0\fn\times\mathbf H_R^{(0)}, 
\end{equation}
while the Rashba field 
\begin{equation}
  \mathbf H_R^{(0)} =  \frac{\dsd}{\hbar v\gamma_0}I_s\alpha_R\fey+\mathcal
O(\alpha_R^3)
\end{equation}
still arises. The first-order contributions essentially remain of the same
form,i.e., 
\begin{align}
 \mathbf T^{\rm{ad}} =& -I_s\dx\fn+\mathcal O(\alpha_R^2)\\
 \mathbf T^{\rm{non-ad}} =& \beta I_s\fn\times\dx\fn+\mathcal O(\alpha_R^2) \,
. 
\end{align}
Only the first-order (nonadiabatic) Rashba field 
\begin{equation}
   \mathbf H_R^{(1)}=
\beta\frac{I_s}{\gamma_0}\alpha_R^2(\dx\fn\cdot\fey)\fey+\mathcal O(\alpha_R^3)\
.
\end{equation}
is now proportional to $\dx\fn\cdot\fey$ and to $\beta\alpha_R^2$. As before, 
this term is only important for very steep DWs, which are not considered in
this work here.

\subsection{Domain wall chirality}

Bloch domain walls have additional degree of freedom which is the chirality. 
 Chirality is defined as the clockwise ($C=+1$) or the
counter-clockwise ($C=-1$) rotation of the magnetic moment in the according
plane. For the system addressed below, an initial direction of the
magnetic moment $n_{x(y)}>0$ at the DW center for the hard $y$ ($x$) axis means
a negative chirality and vice 
versa (cf. Fig. \ref{fig::scheme}). The chirality-dependent DW
dynamics\cite{otalora2013breaking} has already been investigated previously  
for fixed chiralities\cite{linder2013chirality}. In contrast to that, we
here allow the magnetization to dynamically tilt and also to eventually switch
the chirality.

\section{Results}

In 1D systems, we theoretically have the freedom to choose the
directions of the easy and hard axes, the direction of the Rashba-induced field
$\mathbf H_R$
and the chirality of the DW. We will consider systems which always have the easy
axis in the $z$ direction, while the Rashba-induced field points in the $y$
direction. At 
the respective ends of the wire, we enforce $n_z(x\to\pm\infty)=\pm1$ as
boundary conditions to solve the Landau-Lifshitz-Gilbert equation numerically.
 A crucial
point of this work is the effect of the direction of the hard axis and the
initial chirality of the DW on its dynamics. We will show results for four types
of DWs: the hard axis in $x$ or $y$ direction and a positive or negative
chirality $C=\pm 1$.

We choose model parameters which correspond to the estimated
values\cite{miron2011} of Pt/Co/AlO$_x$. We have: $A_{ex}=10^{-11}\rm{J/m}$,
$M_s=1090\rm{kA/m}$, $K_{\parallel}=0.92\rm{T}$, $K_{\perp}=0.03K_{\parallel}$,
$2\alpha=\beta=0.12$. In addition, we set $\dsd=0.5\rm{eV}$, $v_F=10^6\rm{m/s}$
and the polarity of the spin current $P=1$. The values of the Rashba interaction
will be chosen around $\tilde\alpha_R=10^{-10}\rm{eVm}$ which corresponds to
$\Delta_R=0.1\rm{eV}$ and $\alpha_R=\Delta_R/\dsd=0.2$. The DW center $\xdw$ is
defined by the condition $n_z(\xdw)=0$ and the DW
velocity is calculated to be $\vdw=\dt \xdw$.

\subsection{Domain wall velocity}

\begin{figure*}
 \includegraphics[width=\linewidth]{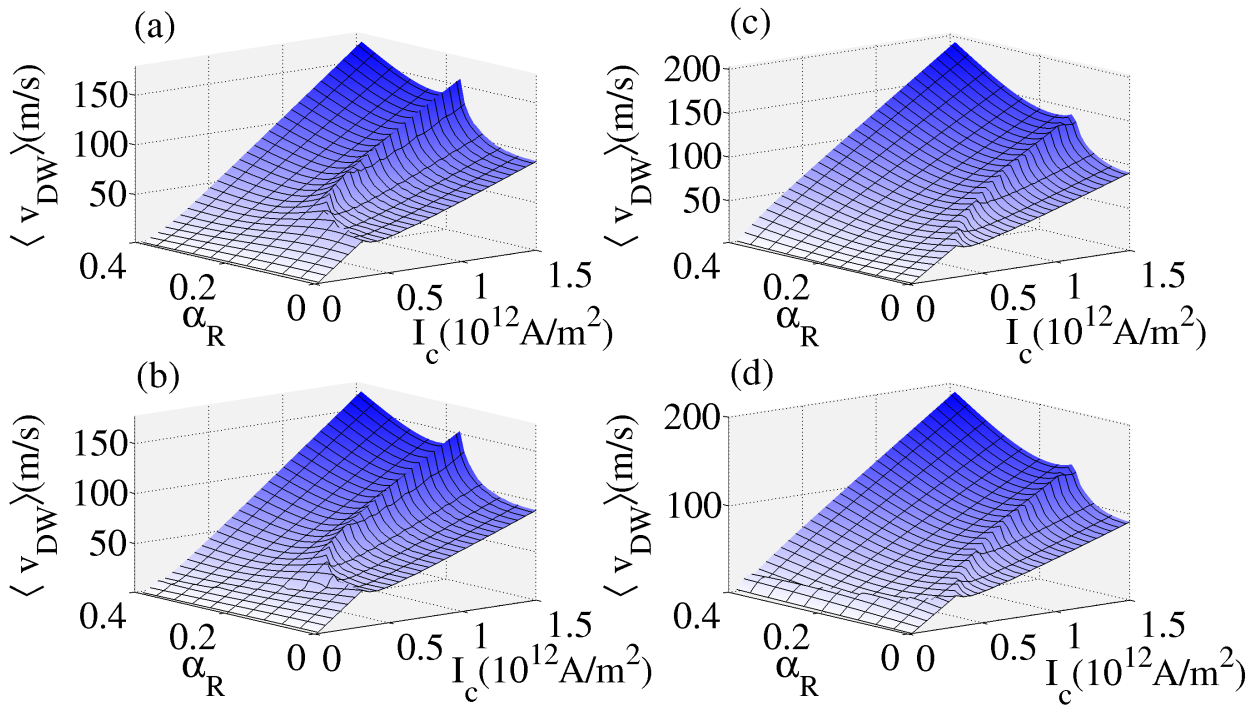} 
 \caption{\label{fig::pd_rel2mp}(color online) Averaged DW velocity for (a) hard
$y$ axis and negative chirality $C=-1$, (b) hard $y$ axis and positive
chirality $C=+1$,(c)
hard $x$ axis and negative chirality $C=-1$, and (d) hard $x$ axis and positive
chirality $C=+1$. Blue colors
indicate positive velocities and red colors negatives ones. The Walker breakdown
occurs at larger current densities for  larger values of the Rashba parameter
$\alpha_R$. The relaxation is assumed to occur towards
$\fJp^{\rm{relax}}\propto\fmp$.}
 \end{figure*}
The DW velocity here depends on several quantities.
First of all, it will be strongly determined by the size of the current
density $I_c$ as well as by the strength of the Rashba spin-orbit interaction
$\alpha_R$. 
Second, there are the topological features which influence the DW dynamics and
which are determined by the sign of the initial
chirality ($C=\pm 1$) and the direction of the hard axis (in $x$ or $y$
direction). Finally, we address the role of the different forms of the
relaxation state
($\fJp^{\rm{relax}}\propto\fn,\fmp$) which effectively decides whether 
 the anti-damping Rashba field $\mathbf H_R^{\rm{anti}}$ appears or not. We 
show the dependence of the DW velocity for all of these cases in this section.

In Fig. \ref{fig::pd_rel2mp}, the long-time  averaged DW velocity $\mean{v_{\rm{DW}}}=x(t_{\rm av})/t_{\rm av}$
, with $t_{\rm av}=100\rm{ns}$,  is
shown for the case of $\fJp^{\rm{relax}}\propto\fmp$, such that $\mathbf
H_R^{\rm{anti}}=0$. All four configurations ($C=\pm 1$, hard $x$ or $y$ axis)
yield a qualitatively similar picture. As we have chosen
$\beta=2\alpha$, we find a Walker breakdown which refers to the rather
sharp velocity drop at small $\alpha_R$ in the vicinity of
$I_c\approx0.3\times10^{12}\rm{A/m^2}$. The
WB is accompanied by a precession of the magnetization at the DW center. For
larger $\alpha_R$, the WB appears at larger current densities. Hence, 
the Rashba field $\mathbf H_R$ stabilizes the DW motion since it suppresses the 
precession of the DW as it acts as an additional effective anisotropy. The same
observation has already been made previously\cite{miron2011,linder2014wb}. To
obtain a rough estimate of the
critical current density $I_c^{(WB)}$, where the WB sets in, we introduce the
total anisotropy field $K^* = K_{\perp}+H_R$. The critical current density is
proportional to the total anisotropy field according to 
\begin{align}
 I_c^{(WB)} =& \nu K^*\\
 =& I_c^{(WB)}(\alpha_R = 0) \left(1 + \frac{H_R}{K_{\perp}}\right) \, .
\end{align}
The remaining proportionality factor can be determined by the critical current
density for a  vanishing Rashba field. In our case, the Rashba field is given
as $H_R\approx\frac{\dsd}{\hbar v_F \gamma_0}\alpha_RI_s=0.25\alpha_R
I_C[10^{12}\textrm{A/m}^2] \textrm{T}$. Thus, upon setting  
$I_c^{(WB)}(\alpha_R=0)=0.3\times10^{12}\rm{A/m^2}$,  we find approximately
that 
\begin{equation}
 I_c^{(WB)} = \frac{0.3}{1-3\alpha_R} \, .
\end{equation}
This yields to a complete suppression\cite{linder2014wb} of the WB for
$\alpha_R\gtrsim 0.3$, which is reflected in Fig. \ref{fig::pd_rel2mp}. 
The differences between the four configurations shown in Fig.
\ref{fig::pd_rel2mp} are discussed in more details in the next section.

\begin{figure*}
 \includegraphics[width=\linewidth]{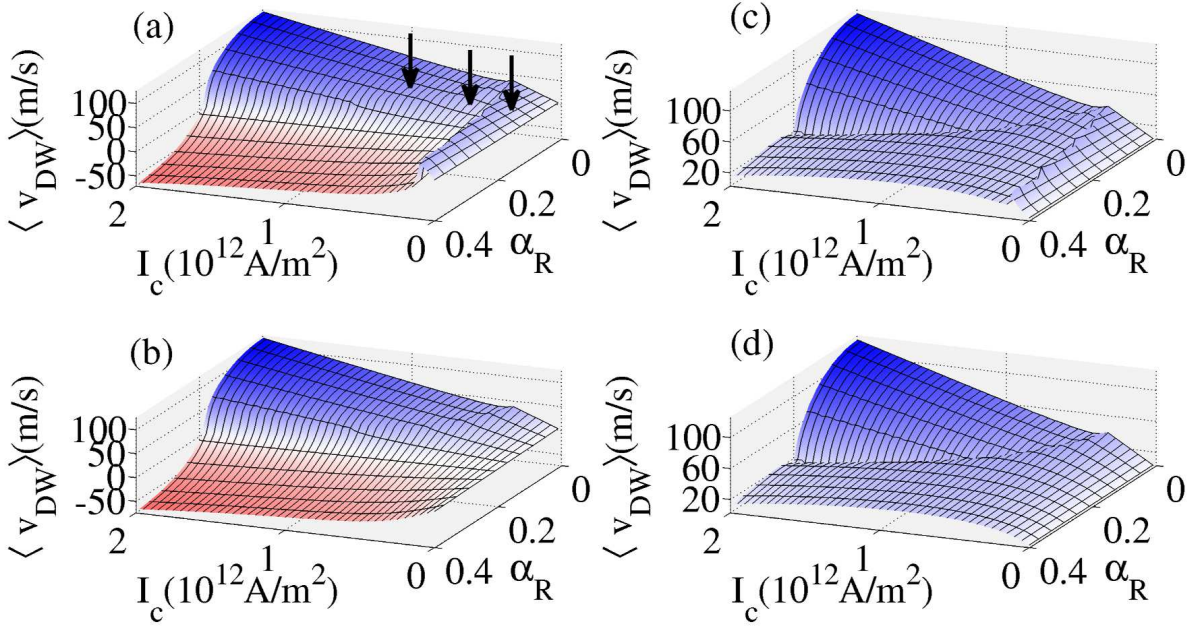}
  \caption{\label{fig::pd_rel2n}(color online) Averaged DW velocity for (a) hard
$y$ axis and negative chirality $C=-1$, (b) hard $y$ axis and positive
chirality $C=+1$, (c)
hard $x$ axis and negative chirality $C=-1$, and (d) hard $x$ axis and positive
chirality $C=+1$. Here, the relaxation occurs towards
$\fJp^{\rm{relax}}\propto\fn$. Blue colors
indicate positive velocities and red colors negatives ones. The black arrows
in (a) mark the cases of $\{I_c,\alpha_R\}$ shown in Fig. \ref{fig::v_t}. }
\end{figure*}

The second case when the relaxation state is $\fJp^{\rm{relax}}\propto\fn$
yields an additional field-like (nonadiabatic) torque with the
anti-damping field $\mathbf
H_R^{\rm{anti}}= -\frac{\dsd}{\hbar v\gamma_0}\beta I_s\alpha_R\fn\times\fey$, 
which is perpendicular to $\mathbf H_R$. The most significant consequence of
the anti-damping term is a possible movement of the DW against the current
flow. This can be seen in Fig. \ref{fig::pd_rel2n} and has already
been discussed in Ref.\ \onlinecite{linder2013chirality}. Again, we devote the
next section to a more detailed discussion of the
differences  between the four configurations. 

\subsection{Chirality switching}

\begin{figure}
 \includegraphics[width=\linewidth]{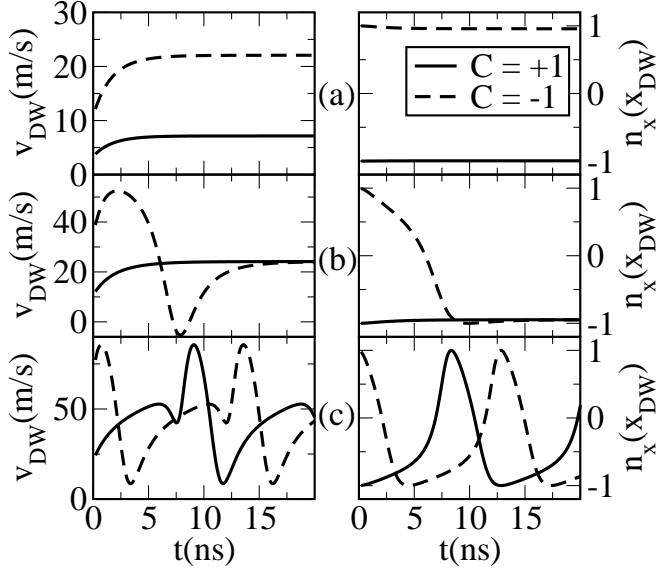}
 \caption{\label{fig::v_t}DW velocities (left) and magnetization in
$x$-direction at DW center (right) for three different current densities: (a)
$I_c=0.125\times10^{12}$A/m$^2$, (b) $I_c=0.4\times10^{12}$A/m$^2$ and (c)
$I_c=0.8\times10^{12}$A/m$^2$. The sign of the $n_x(x_{DW})$ indicates the
(inverse) chirality of the DW. Three scenarios appear: (a) no switching of the
chirality, (b) a single chirality switching for a positive initial chirality,
and, (c) alternating chirality switching (and a Walker breakdown). Parameters
used are $\alpha_R=0.1$, hard $y$
axis, and the relaxation to $\fJp^{\rm{relax}}\propto\fn$.}
\end{figure}

Even though the DW velocities in Figures \ref{fig::pd_rel2mp} and
\ref{fig::pd_rel2n} show a qualitatively similar behavior for all four
configurations, several differences between them arise. They become more
explicit when we investigate the time evolution
of the DW velocity in more detail. Figure \ref{fig::v_t} shows the dynamical
build-up of the DW velocity for the
three parameter sets $\{\alpha_R,I_c\}$ indicated by the black arrows in Fig.
\ref{fig::pd_rel2n}. In this case, the hard axis is in the $y$ direction and
the results are shown for both initial chiralities $C=\pm 1$.
Three scenarios can be identified: (a) For low current densities $I_c$, the two
chiralities lead to different DW velocities for all times. (b) 
For intermediate $I_c$, the initially different velocities approach each
other after some time. (c) Finally, for large current densities $I_c$, the two
chiralities lead to equal but phase-shifted oscillating velocities with the
same rather large average velocity. We also show in Fig. \ref{fig::v_t} the
magnetization $n_x(x_{\rm{DW}})$ in the  $x$ direction at the DW center. It is 
immediately clear what separates these three scenarios. In the case (a), the
initial magnetizations of both chiralities remain unchanged over time and
 each chirality is conserved. In this case, neither the Rashba field $H_R\propto
\alpha_R I_c$ nor the
non-adiabatic torque $T^{\rm{non-ad}}\propto\beta I_c$ are strong enough to
overcome the field of the perpendicular anisotropy $K_{\perp}$. In the 
scenario (b), the Rashba field is stronger than $K_{\perp}$.
In contrast to the field of the anisotropy $H_{\perp} = K_{\perp} n_{\perp}$,
the Rashba field explicitly favors one direction of the magnetization
$n_x(x_{\rm{DW}})\gtrless0$. For the case shown in Fig. \ref{fig::v_t} (b), a
negative magnetization is preferred and an initially negative chirality is
switched to a positive one after some time. Finally, for large current
densities and scenario (c), the non-adiabatic torque may overcome both the
perpendicular anisotropy field and the Rashba field. This yields, as usual, to a
Walker breakdown. Here, both  chiralities are alternatingly switched and the
magnetization at the DW center precesses around the $z$ axis.

\begin{figure}  
\includegraphics[width=\linewidth]{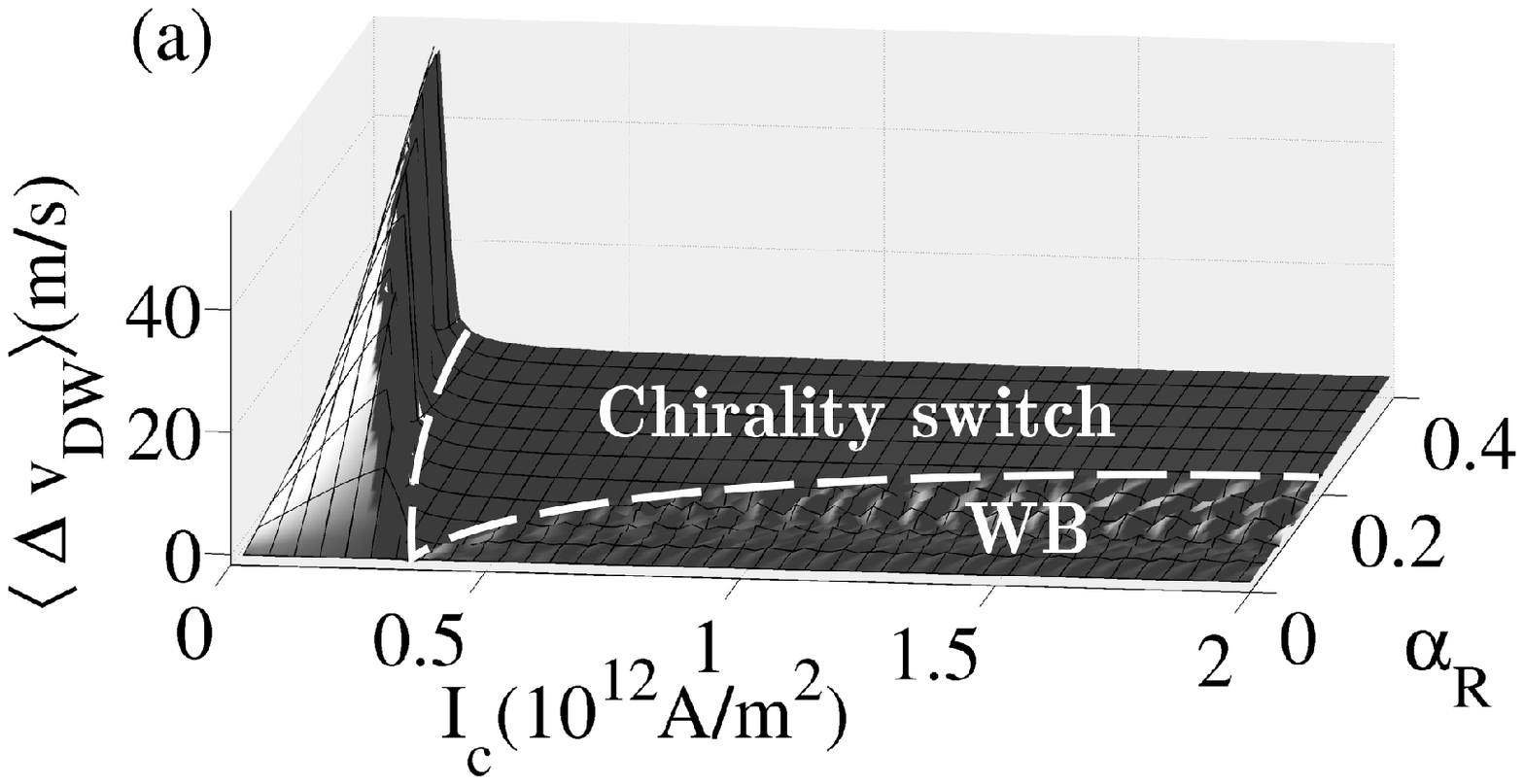}
 \includegraphics[width=\linewidth]{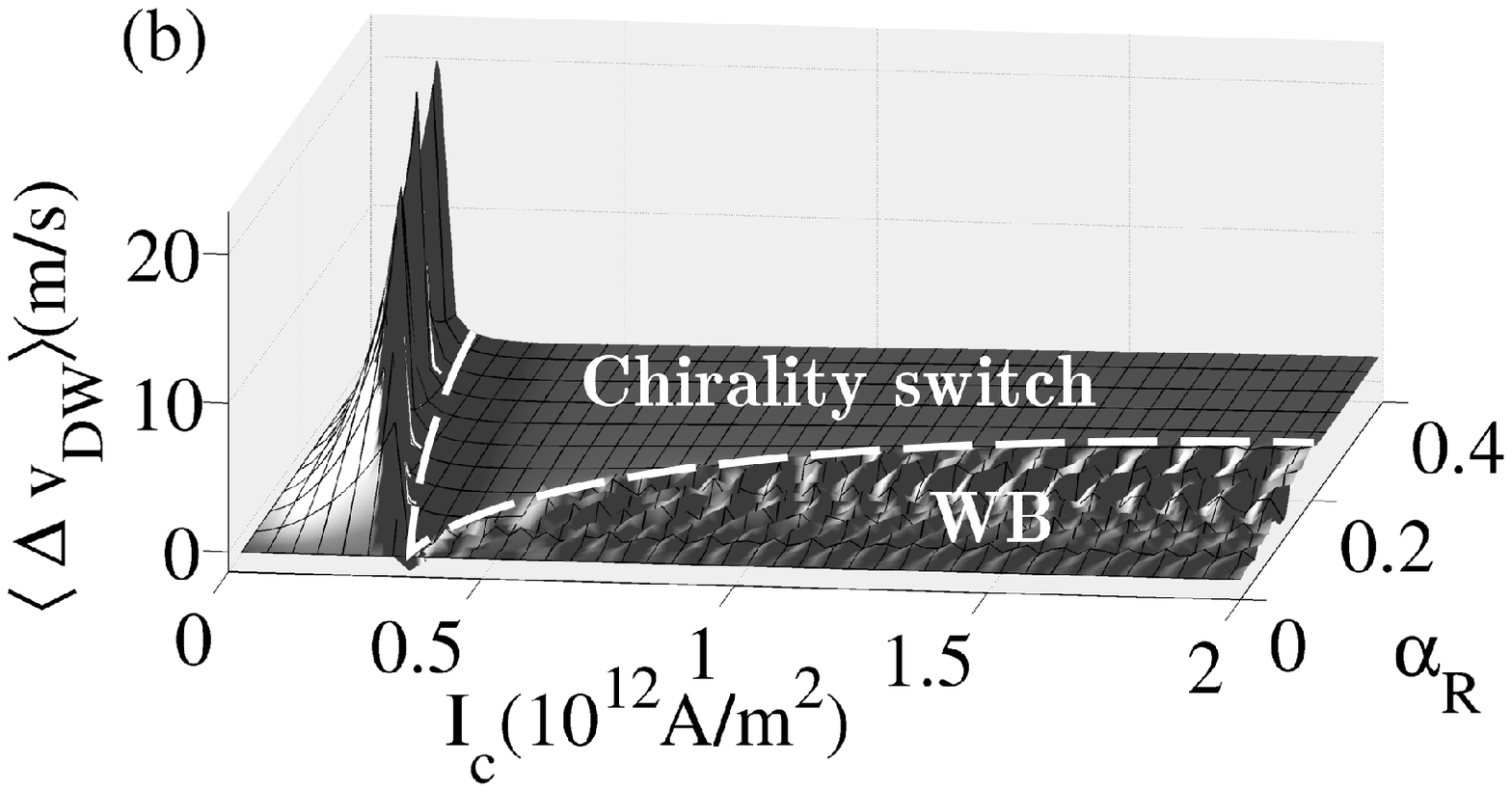}
 \caption{\label{fig::diffv_rel2n}Difference between the DW velocities
$\mean{\Delta v_{DW}}=\mean{v_{DW}^{C=-1}}-\mean{v_{DW}^{C=+1}}$ of both
chiralities for (a) hard $y$ axis and (b) hard $x$ axis. White dashed lines
separate the three scenarios shown in Fig. \ref{fig::v_t}. The relaxation occurs
towards $\fJp^{\rm{relax}}\propto\fn$.}
\end{figure}
\begin{figure}  
\includegraphics[width=\linewidth]{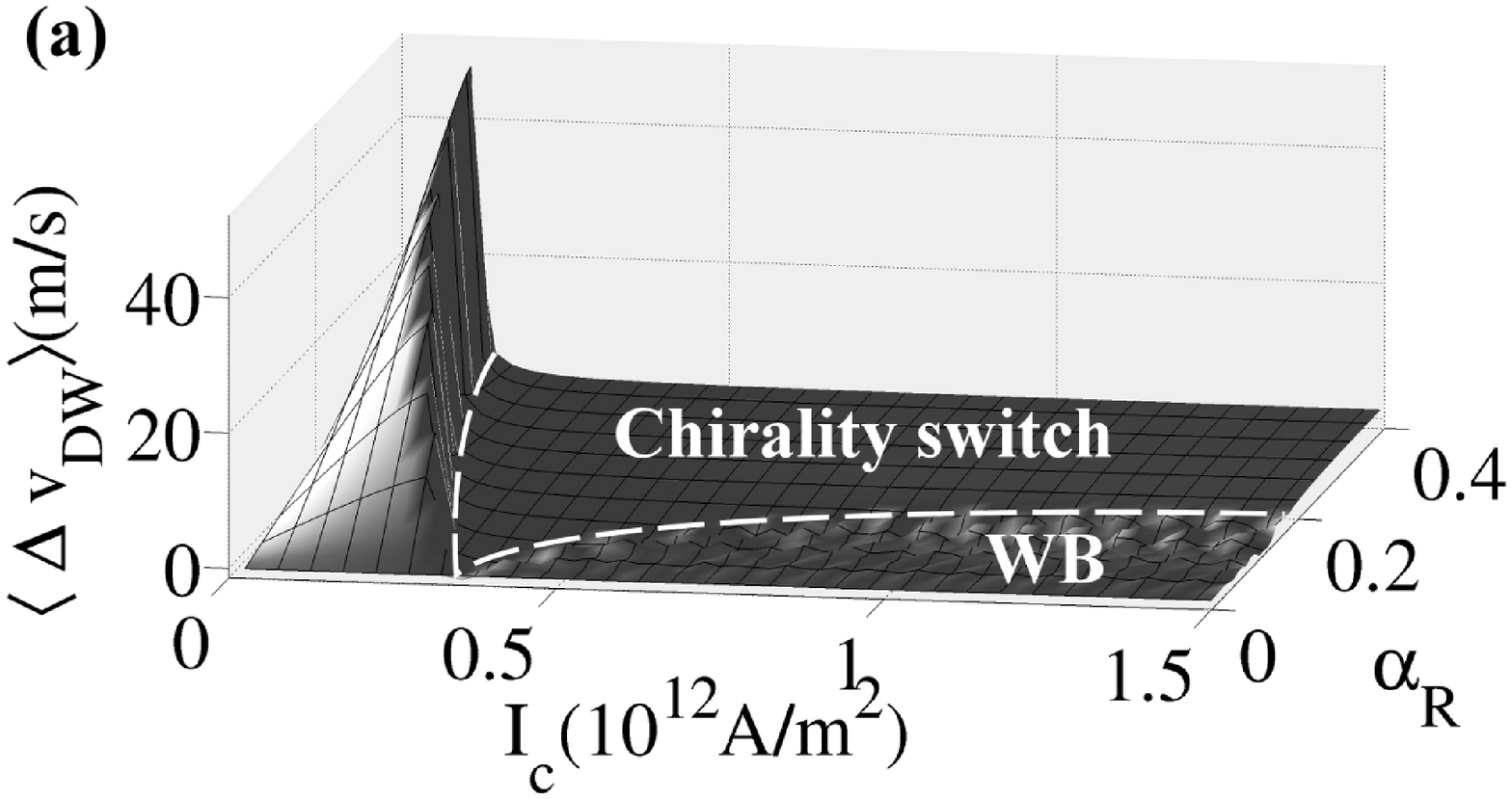}
 \includegraphics[width=\linewidth]{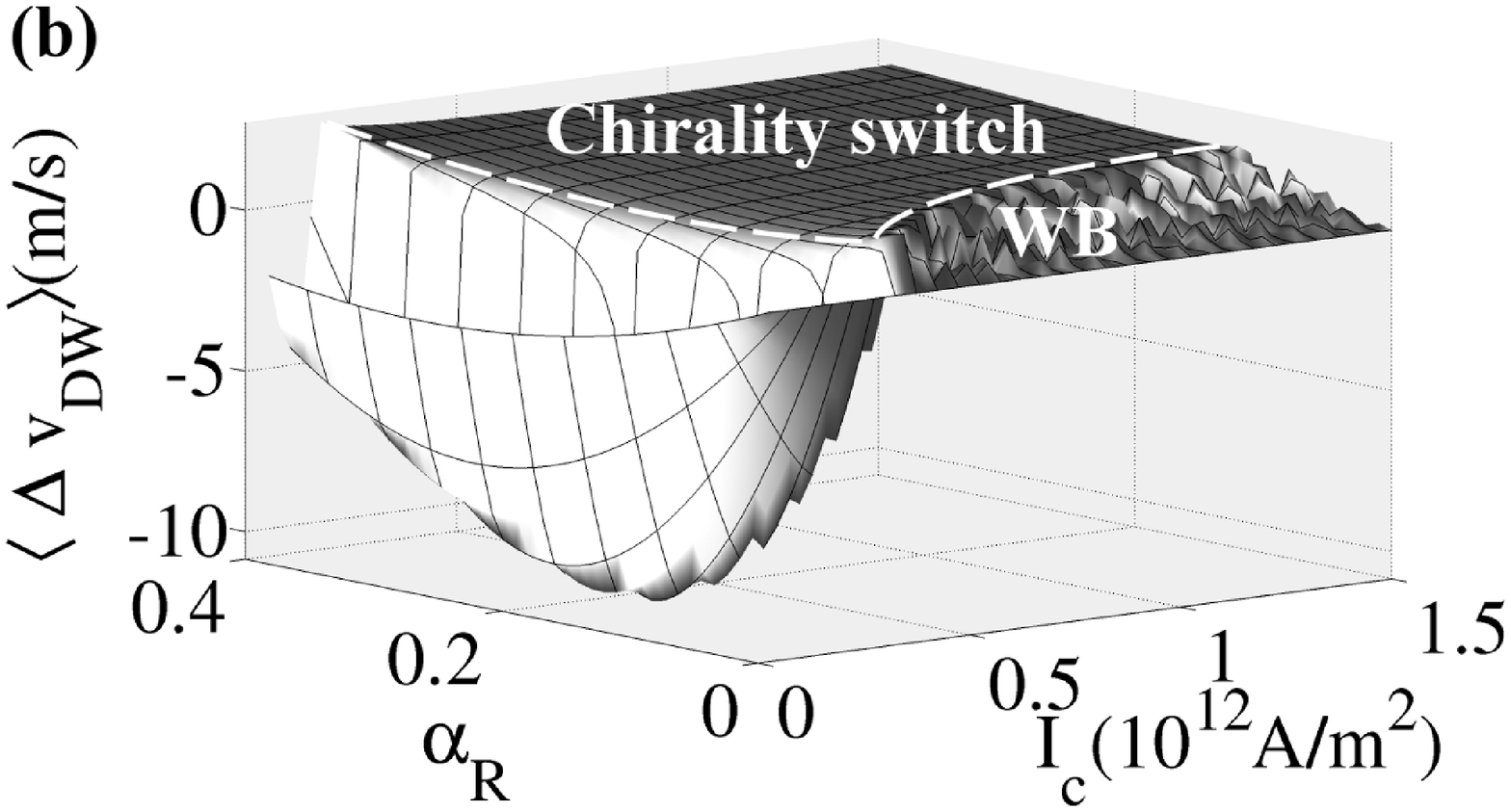}
 \caption{\label{fig::diffv_rel2mp}Same as in Fig. \ref{fig::diffv_rel2n}, but
the relaxation occurs towards $\fJp^{\rm{relax}}\propto\fmp$. }
\end{figure}

The conditions for the three scenarios are that in (a), there
is no chirality switching if $H_R,T^{\rm{non-ad}}<K_{\perp}$, in (b) a
\textit{single}
chirality switching to a preferred chirality occurs if
$T^{\rm{non-ad}},K_{\perp}<H_R$, and in (c), a frequent alternating chirality
switching of both initial chiralities and a WB arise, if
$H_R,K_{\perp}<T^{\rm{non-ad}}$.

We may summarize these scenarios in a phase diagram. For this, we
have calculated the difference $\mean{\Delta
v_{\rm{DW}}}=\mean{v_{DW}^{C=-1}}-\mean{v_{DW}^{C=+1}}$ of the averaged 
velocities of both chirality classes. The results are shown
in Figs. \ref{fig::diffv_rel2n} and \ref{fig::diffv_rel2mp}. For small current
densities in scenario (a), large velocity differences arise and no chirality
switching occurs. Both chiralities yield to permanently different velocities. 
Instead, for intermediate/high current densities, a
single chirality switching occurs and the velocity differences are almost
vanishing. Furthermore, at larger current densities but small $\alpha_R$, 
we find small oscillations of $\mean{\Delta v_{\rm{DW}}}$ which stem from the
 alternating chirality switchings in scenario (c). Even though we have
performed a long-time average, the time interval $t\in[0,100\rm{ns}]$ over
which the DW velocity is averaged, is not long enough to completely
remove these oscillations. However, even though we could increase the time
window of the average, we prefer to use these oscillations for an easier
determination of the WB ``phase''. 

Finally, we note that in this work, we have focused on the case $\beta>\alpha$ 
which is a necessary condition for a WB to appear. For the case $\alpha=\beta$, 
the ``phase'' of the WB would vanish. In addition, in order to achieve a
chirality switching, larger values of $\alpha_R$ or $I_c$ would be required. 
This
is because the non-adiabatic torque assists the chirality switching in the same
way as it already tends to switch the chirality frequently in the form of a
WB.

To summarize this section, we have shown that it is in principle possible
to control the chirality of a DW by the size of the applied current and that
clearly separated phases arise which render this control achievable. This
feature could be useful in technological applications in form of magnetic
storage devices. Even though we have shown here only results where the switching
to one preferred chirality is illustrated, it could
be easily modified by the direction of the current flow. As $H_R\propto I_c$, 
a current applied in the opposite direction would change the sign of the
Rashba field, which would then favor the other chirality. 

\subsection{Short current pulses}

\begin{figure}
 \includegraphics[width=\linewidth]{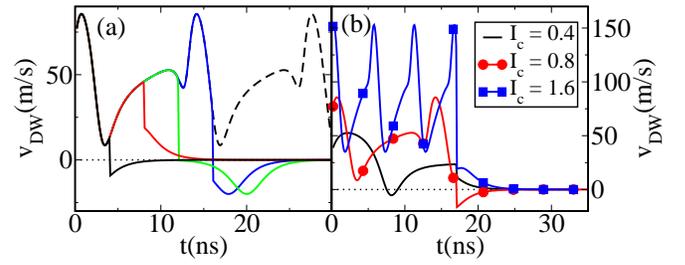}
 \caption{\label{fig::v_vs_t_pulse}(color online) DW velocity vs time for (a)
different current pulses with pulse lengths $t_p=\{4,8,12,16\}\rm{ ns}$ and
the steady current
(dashed line) at $I_c=0.8\times 10^{12}\rm{A/m^2}$ and (b) different current
densities $I_c=\{0.4,0.8,1.6\} \times 10^{12}\rm{A/m^2}$ for $t_p=17\rm{ns}$.
After the current pulse ends, the DW drifts some distance in either positive or
negative direction depending on the state of the DW at the end of the pulse.
Parameters: $\alpha_R=0.1$, hard $y$ axis, negative chirality $C=-1$.}
\end{figure}

Most of the results for the DW velocities in the previous section refer to
averages over a intermediate-to-large time window. However, as shown in
Fig. \ref{fig::v_t}, relevant features arise on much shorter time scales, e.g., 
in the regime of a few
nanoseconds\cite{stier2013,thomas2006oscillatory,meier2007direct}. For example a
chirality switching as in Fig. \ref{fig::v_t} (b) only appears after some
finite time. The long-time averaged velocity shows no difference, 
because for most of the time, both velocities match each other. However, an 
analysis in form of a shorter-time average could uncover the different
velocities of the DWs of different initial chirality. 

In this section, we address the DW dynamics on shorter time scales and use for
this rather short current pulses, which basically includes
short time averaging. It is known that several new features appear for short
current pulses\cite{stier2013,miron2011,thomas2010dynamics}. First, possible
oscillations, which would build up on longer times, will not average out for 
pulse lengths of the order of the oscillation period. This can readily be seen
in Fig. \ref{fig::v_vs_t_pulse}. When a pulse ends, e.g., in the first half of
an 
oscillation period, it can yield a drastically increased or decreased
DW velocity, depending on the sign of the amplitude in the
respective half-period. This should also lead
to major chirality-dependent velocity differences since the oscillation
amplitudes may differ for both chiralities [cf. Fig. \ref{fig::v_t}(c)].

Second, the DW does not immediately stop after the end of the current pulse, but
``drifts'' a certain distance either in forward or backward
direction\cite{stier2013,miron2011,thomas2010dynamics}. The 
direction of the drift is determined by the momentary state of the DW at the
end of the pulse. As there are no spin torques any more acting after the pulse  
(neither any Rashba fields nor the conventional (non-)adiabatic torques are
present), the DW
strives to settle at its equilibrium position which is determined by the hard
axis anisotropy. If the DW is tilted out of this position at the end of the
pulse, it realigns in the fastest manner back to it and thereby moves some
distance.

Not only the pulse length $t_p$ determines the state of the DW at the end of
the pulse, but also the current density. For a constant $t_p$, an increasing
$I_c$ leads to a smaller oscillation period as it is shown in Fig.
\ref{fig::v_vs_t_pulse} (b). Thus the state at the end of the pulse, and with it
the drift, is changed.

To see the effects of the short-time averaging and the drifting, we
define two different averaged velocities. First, the averaged velocity during
the pulse is 
$\mean{v_{\rm{DW}}}=x_{\rm{DW}}(t_p)/t_p$, and, second, an effective averaged
velocity $\mean{v^{\rm{eff}}_{\rm{DW}}}=x_{\rm{DW}}(t\to\infty)/t_p$ is
meaningful. The second quantity includes the drifting since we use the DW
position $x_{\rm{DW}}(t\to\infty)$ after a long enough time. As it is difficult
to decide at which time the DW actually
stops, we also divide this position by the pulse length $t_p$. Thus, the
effective velocity $\mean{v^{\rm{eff}}_{\rm{DW}}}$ is not a real time average,
but an ``effective'' one which is compared to the pulse length.

\begin{figure}  
\includegraphics[width=\linewidth]{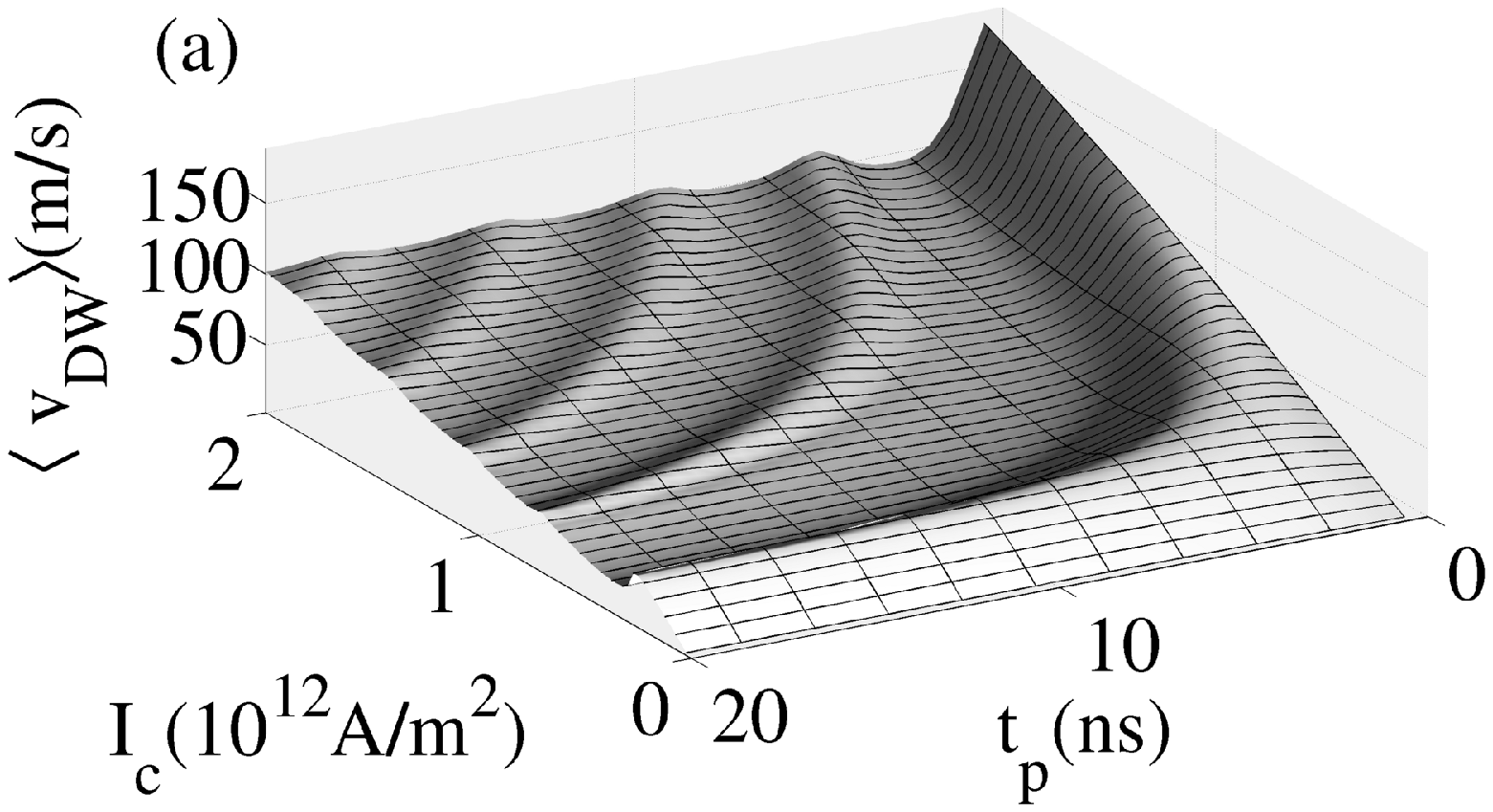}
 \includegraphics[width=\linewidth]{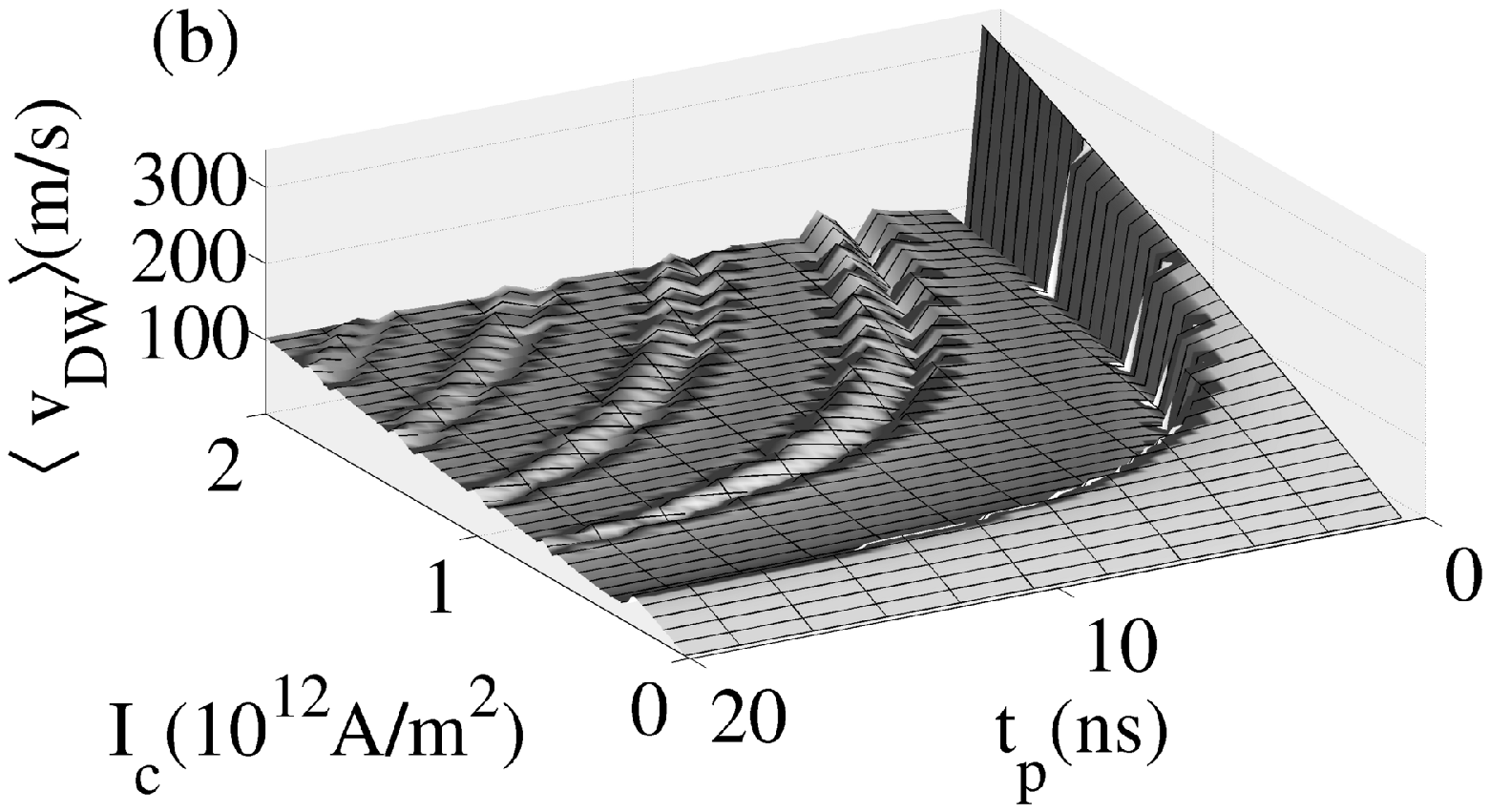}
  \includegraphics[width=\linewidth]{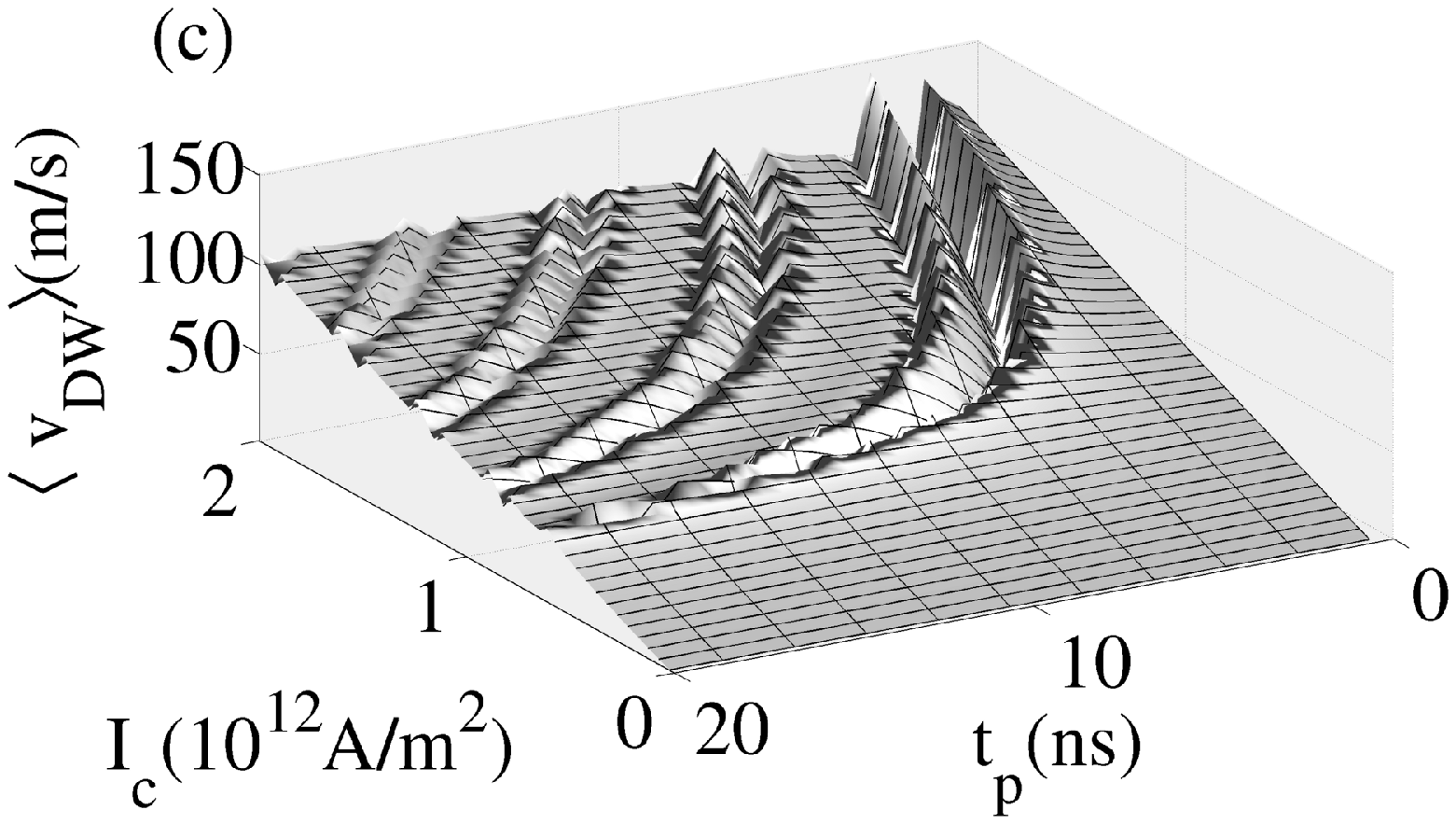}
 \caption{\label{fig::v_vs_pulse}(a) Average velocity over the current pulse
length $\mean{v_{DW}}=x_p(t_p)/t_p$ and (b) effective velocity
$\mean{v_{DW}}=x(t\to\infty)/t_p$ which includes the drifting of the DW after
the pulse has ended, both for a hard $y$ axis and negative chirality $C=-1$. 
While the average velocity in (a) shows smooth oscillations at higher current
densities due to the Walker breakdown, the drifting after the pulse leads to
rather abrupt velocity changes.
(c) Same as (b) but for positive chirality $C=+1$. Shown are the
results for $\alpha_R=0.1$ and for the relaxation to
$\fJp^{\rm{relax}}\propto\fn$.}
\end{figure}

Figure \ref{fig::v_vs_pulse} shows the DW velocities versus the
current density $I_c$ and the pulse duration $t_p$. We find smooth
oscillations of the average velocity $\mean{v_{DW}}$, i.e., without
drifting. They occur because the oscillatory motion of the DW does not completely average out 
even for large current densities. For very short pulses, remarkably large 
velocities appear for a DW with negative chirality which can be traced back
to the large velocities at the beginning of the motion for this chirality in
systems with a hard $y$ axis [cf. Fig. \ref{fig::v_t} (c)]. 
When we include the drifting and consider the effective mean velocity 
$\mean{v^{\rm{eff}}_{\rm{DW}}}=x_{\rm{DW}}(t\to\infty)/t_p$, the
changes in the velocity become more abrupt [cf. Fig. \ref{fig::v_vs_pulse} (b)].
Then, the drifting partially compensates the oscillatory movement to some extent
due its different moving directions. This lets the DW end up in distinct
positions. 

\begin{figure}  
\includegraphics[width=\linewidth]{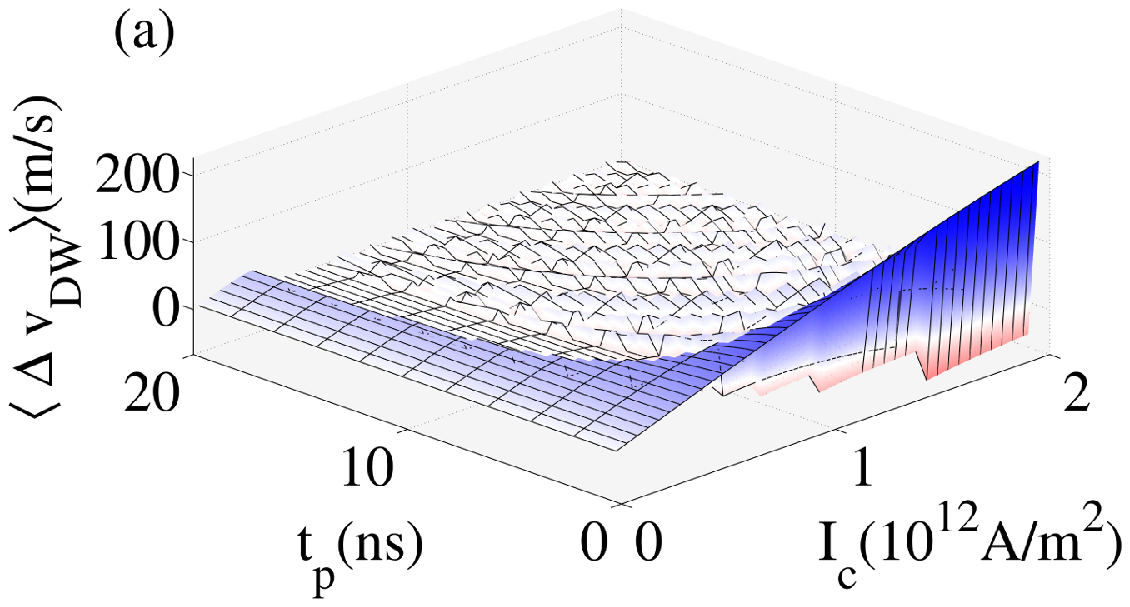}
 \includegraphics[width=\linewidth]{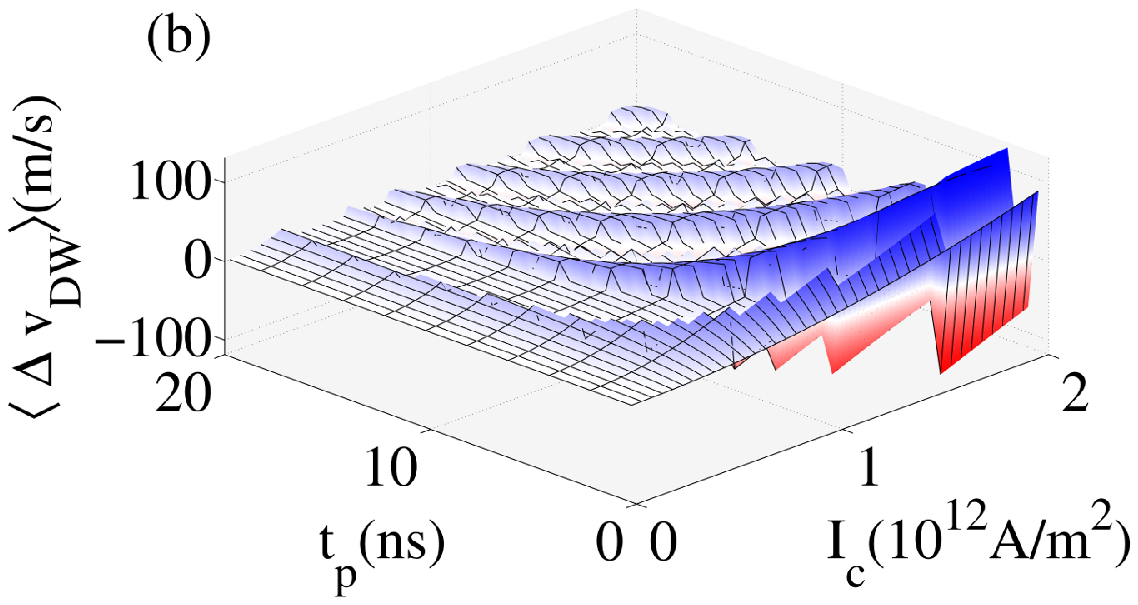}
 \caption{\label{fig::diffv_vs_pulse}(color online) Differences of effective
velocities $\mean{\Delta
v_{DW}}=[x^{C=-1}(t\to\infty)-x^{C=+1}(t\to\infty)]/t_p$ for DWs with (a) a hard
$y$ axis, and, (b) a hard $x$ axis. Particularly for short pulses, large
differences occur. The blue color indicates positive and red negative values.
Moreover, $\alpha_R=0.1$, and the relaxation occurs to
$\fJp^{\rm{relax}}\propto\fn$.}
\end{figure}

In contrast to the rather large velocities of the DW with negative chirality,
a positive chirality $C=+1$ leads to a strongly reduced DW velocity at very
short pulses [cf. Fig. \ref{fig::v_vs_pulse} (c)]. Thus,  large velocity
differences between the two cases of opposite initial chiralities at small
$t_p$ are expected. Figure \ref{fig::diffv_vs_pulse} indeed confirms this.
 In addition to the large absolute velocity differences at small $t_p$, these
differences also vary very strongly and can actually change the sign and with
that the motional direction of the DW. 

These results illustrate the important role of the additional short-time
effects for the DW dynamics, in particular in the presence of a Rashba
spin-orbit interaction. Its impact on specific experiments may also be affected
by additional phenomena not considered in this work, such as the presence of
pinning centers\cite{van2013role}, for instance. 

\section{Summary}

We have studied the influence of the direction of the hard axis
and the DW chirality on the DW's current induced dynamics in 1D Rashba wires.
The spin transfer torque arises from a gradient expansion in the DW steepness
and enters in the Landau-Lifshitz-Gilbert equation which we solve numerically.
Two different relaxation states were used which either generate or suppress 
the  Rashba anti-damping field-like torque $\mathbf H_R^{\rm{anti}}$. This is 
perpendicular to the
regular Rashba-induced field-like torque $\mathbf H_R$, which arises in both
cases.

As we focus on the case when the nonadiabaticity parameter $\beta > \alpha$,
a Walker breakdown at sufficiently strong current densities $I_c$ arises. It 
 is suppressed by the Rashba field $\mathbf H_R$, since it acts as an
additional local anisotropy. For rather strong Rashba couplings $\alpha_R$, 
the WB is entirely suppressed. Even more interestingly, we identify a third
``phase'' in the $\{\alpha_R,I_c\}$ parameter space: For intermediate current
densities and/or large $\alpha_R$, we find a single switching of the initial
chirality to a preferred chirality, which can be chosen by the direction of the
current flow. 

Moreover, the effects of the chirality switching appear to be more pronounced
at short times and we have considered short current pulses. As expected, we
find a stronger influence of the momentary velocity state when a short-time 
averaging procedure is applied. 
Further phenomena such as a drifting of the DW after a short current pulse
affect the short-time dynamics of a DW even more pronounced. This results partly
in rather abrupt changes of the effective DW velocity and  a
completely different dependence on the current density as compared to the steady
current arises. Then, even larger velocity differences between DWs of
different chiralities result. This rich set of features shows that several
possibilities arise for optimal parameter combinations in order to achieve a
maximal DW velocity. 

\begin{acknowledgments}
We acknowledge support from the DFG SFB 668 (project B16) and thank
G. Meier for valuable discussions.
\end{acknowledgments}

%

\end{document}